\documentclass[journal]{IEEEtran}

\usepackage[T1]{fontenc}
\usepackage[utf8]{inputenc}
\usepackage{cite}
\usepackage{graphicx}
\usepackage{amsmath,amssymb,amsfonts}
\usepackage{siunitx}
\usepackage{booktabs}
\usepackage{url}
\usepackage{hyperref}
\hypersetup{hidelinks}

\usepackage[switch]{lineno}

\title{Fast and bright scintillators for ultrafast materials dynamics using 4$^{th}$ generation synchrotron}

\author{Zhehui Wang,~\IEEEmembership{Senior Member,~IEEE,}
        Dana Dattelbaum, Rachel Huber, Quanxi Jia, Yuelin Li, Katie Liu, C. L. Morris, Brad D. Price, Robert Reinovsky, Adam Schuman, Nicholas Sinclair,
        Maria Spiropulu, Yoshimasa Toyoda, Christina Wang, Li-Yuan Zhang, and~Ren-Yuan Zhu,~\IEEEmembership{Senior Member,~IEEE}%
\thanks{Manuscript received Month DD, 2026; revised Month DD, YYYY; accepted Month DD, YYYY. This work was supported in part by ...}%
\thanks{Zhehui Wang, Dana Dattelbaum, Rachel Huber, C. L. Morris and Robert Reinovsky are with Los Alamos National Laboratory, Los Alamos, NM 87545, USA (e-mail: zwang@lanl.gov).}%
\thanks{Quanxi Jia is with University at Buffalo, Buffalo, New York 14260, USA.}%
\thanks{Yuelin Li is with Argonne National Laboratory, Lemont, IL 60439 and Washington State University, Pullman, WA 99164, USA.}
\thanks{Katie Liu is with Brown University, Providence, RI 02912, USA.}%
\thanks{Brad D. Price, Adam Schuman, Nicholas Sinclair and Yoshimasa Toyoda are with Washington State University, Pullman, WA 99164, USA.}
\thanks{Maria Spiropulu, Li-Yuan Zhang, and~Ren-Yuan Zhu are with California Institute of Technology, Pasadena, CA 91125, USA.}%
\thanks{Christina Wang was with California Institute of Technology, Pasadena, CA 91125, USA, and is currently with Fermi National Accelerator Laboratory (Fermilab), Batavia, IL 60510, USA}%
\thanks{Digital Object Identifier 10.1109/TNS.2026.XXXXXXX}}

\begin{document}
\maketitle

%\linenumbers

\begin{abstract}
We present recent advances in fast and bright scintillators for ultrafast X-ray phase contrast imaging of dynamic materials experiments at the upgraded Advanced Photon Source (APS-U), a fourth (4$^{th}$) generation synchrotron. APS-U enables hard X-ray imaging at frame rates of at least 13 MHz (corresponding to 77 ns or shorter interframe intervals), creating a new need for scintillators with faster response and higher light output than lutetium–yttrium oxyorthosilicate (LYSO). For indirect imaging and diffraction with ultrafast cameras, commercial lanthanum bromide (LaBr\textsubscript{3}) and cerium bromide (CeBr\textsubscript{3}) are promising candidates. These materials exhibit decay times approximately a factor of two shorter than LYSO ($\sim$ 40 ns) and lutetium oxyorthosilicate (LSO), while maintaining comparable light yield per incident X-ray photon. However, their implementation at APS-U requires addressing several challenges, including material limitations due to hygroscopicity, efficient optical coupling to imaging systems, and high quantum efficiency for conversion of scintillation light, predominantly at wavelengths below 400 nm, into detectable electronic signals. We report results from material characterization, detector integration and packaging, and beamline experiments of materials with impact. In addition, emerging scintillator classes, including perovskites and high-entropy materials, are discussed as potential alternatives for next-generation ultrafast X-ray diagnostics.
\end{abstract}

\begin{IEEEkeywords}
Ultrafast X-ray phase contrast imaging, hygroscopic effects, cerium-doped scintillators, 4th generation synchrotron, hermetic sealing.
\end{IEEEkeywords}

\section{Introduction}

The advent of fourth-generation synchrotron light sources and other high-energy X-ray sources  has fundamentally transformed the ability to probe matter under extreme and dynamic conditions. Facilities such as the upgraded Advanced Photon Source (APS-U) now enable high-brightness, high-energy X-ray beams with repetition rates above 10 megahertz, opening a pathway to real-time ``movies'' of mesoscale structural evolution in materials. These capabilities are central to a wide range of scientific and technological-driven applications, including phase transformations, shock-compressed matter, additive manufacturing, and radiation effects in nuclear materials.

However, ultrafast detector technology has emerged as the dominant bottleneck limiting scientific exploitation of these sources. While modern hybrid pixel detectors and direct detection systems have achieved frame rates on the order of $\sim$10~MHz at moderate photon energies, they fall significantly short of the requirements imposed by next-generation facilities, which demand frame rates exceeding 10--100~MHz, sensitivity and high detection efficiency to photon energies above 20~keV, and high dynamic range within a single pulse sequence. As highlighted in recent community assessments, there exists a substantial performance gap between the current state-of-the-art and the requirements for ultrafast, high-energy X-ray imaging \cite{Wang2017ultrafast}.

Indirect detection approaches based on scintillators remain indispensable for many experimental configurations, particularly where large-area imaging, flexible optical coupling, or compatibility with ultrafast visible-light cameras is required. In this paradigm, scintillators serve as the critical transducer between hard X-ray photons and optical detection systems. Consequently, the performance of the scintillator directly determines the achievable temporal resolution, signal-to-noise ratio, and overall system efficiency.

One of the most widely used scintillators, such as lutetium--yttrium oxyorthosilicate (LYSO), and equivalent elutetium oxyorthosilicate (LSO), provide high X-ray attenuation power and high light yield, but their intrinsic decay times ($\sim$40~ns) impose a fundamental limitation on frame rates about factor 2-3 below below those required for APS-U 48-bunch mode and similar facilities. At frame intervals of 77~ns or shorter (324-bunch mode, the `brightness mode' designed to maximize X-ray brilliance and coherent flux), significant signal pile-up and temporal cross-talk occur, leading to loss of contrast and degradation of quantitative measurements. This limitation is not incremental but fundamental: without a step change in scintillator response time and brightness, ultrafast hard X-ray imaging at next-generation sources cannot be realized.

The challenge is further compounded by competing material and system-level constraints. High-Z materials are required for efficient absorption of hard X-rays, yet increased stopping power typically comes at the expense of slower scintillation response or reduced material stability. Efficient light extraction and optical coupling must be maintained at emission wavelengths often below 400~nm, where detector quantum efficiency is limited. In addition, many promising fast scintillators, such as halide-based crystals, suffer from hygroscopicity and long-term degradation, posing significant challenges for deployment in high-flux beamline environments.

Recent efforts have identified several promising candidate materials, including cerium-doped lanthanum bromide (LaBr$_3$:Ce) and cerium bromide (CeBr$_3$), which offer decay times significantly shorter than LYSO while maintaining comparable light yield~\cite{HZZC:2018,HZZD:2019}. Emerging materials such as BaF$_2$, perovskites, and nanostructured scintillators further suggest the possibility of sub-nanosecond or even picosecond response. However, these materials introduce new challenges in light yield, radiation hardness, manufacturability, and system integration, and none have yet demonstrated the combination of speed, brightness, and robustness required for routine use in ultrafast hard X-ray imaging.

At the system level, the problem is inherently coupled: advances in scintillator materials must be matched by developments in optical coupling, pixelated imaging photodetectors, and high-speed readout electronics. Burst-mode imaging architectures, through the use of temporary storage memories, partially mitigate data bandwidth limitations, but they place even stricter requirements on scintillator response and light output to preserve signal fidelity at high frame rates. As a result, the development of next-generation scintillators is not merely a materials problem, but a system-critical challenge that directly impacts the feasibility of ultrafast X-ray diagnostics.

In this work, we present recent progress in the development and deployment of fast and high-brightness scintillators for ultrafast hard X-ray imaging at APS-U. We focus on commercially available cerium-based scintillators as near-term solutions, addressing key challenges in material stability, packaging, and optical coupling. We report results from laboratory and the DCS/APS-U beam line characterization under realistic operating conditions and evaluate performance relative to established materials such as LYSO. In addition, we assess emerging material platforms, including perovskites and high-entropy scintillators, as potential pathways toward next-generation detectors capable of meeting the stringent requirements of future ultrafast X-ray experiments.

\section{Scintillators and Characterization \label{sec:SC}}

The scintillator samples, primarily LaBr$_3$:Ce and CeBr$_3$, were obtained from various commercial vendors. The samples come in two materials forms, powders and single crystals~\cite{knoll2010}. Alternates, such as perovskites, high entropy thin scintillators, were not available at the time and will be left for future study. Some of the important materials properties for ultrafast X-ray imaging are summarized in Table.~\ref{tab:properties}.

\begin{table*}[!hbt]
\caption{Scintillators and materials properties that are important to ultrafast X-ray imaging. Values are representative room-temperature literature/vendor values; attenuation lengths are approximate 1/e lengths at 25 keV (without coherent scattering).}
\label{tab:properties}
\centering
\begin{tabular}{lcccccc}
\toprule
Material & Density & Decay time & Light yield & Emission Peak & Refractive index & 25-keV X-ray Attenuation length\\
& (g/cm$^{-3}$) & (ns) & (ph/keV) & $\lambda$ (nm) & & ($\mu$m)\\
\midrule
LYSO:Ce      & 7.1-7.25  & 36-45    & 30-33 & 420 & 1.81-1.83 & 50--51\\
LaBr$_3$:Ce  & 5.08 & 16-25    & 60-70 & 380 & 1.90-1.95 & 71--72\\
CeBr$_3$     & 5.1-5.2  & 17--22 & 60-68 & 380 & 2.09-2.1 & 68--70\\
\bottomrule
\end{tabular}
\end{table*}

\subsection{Assembly of Hygroscopic Scintillators \label{sec:AHS}}

After selecting the appropriate scintillator thickness, particularly for the crystal forms, the primary challenge in using LaBr$_3$:Ce and CeBr$_3$ was to prevent hygroscopic degradation during detector construction and use. The scintillators were assembled inside a flowing dry nitrogen glove box according to established procedures. Each scintillator was coupled to top and bottom optical windows, forming a three-layer “sandwich” configuration with the scintillator at the center. The minimal window thickness of 250 \textmu m was a balance of robustness of the assembly and the light transmission path length through the windows. Both windows are transparent, which allow imaging from both sides of the scintillator, a common mode of ultrafast X-ray imaging using multiple intensified cameras at DCS/APS-U (see below for the setup and other details). Several epoxies from different manufacturers were evaluated, producing final assemblies with either black or transparent epoxied edges.

\begin{figure}[!t]
\centering
\includegraphics[width=3.6 in]{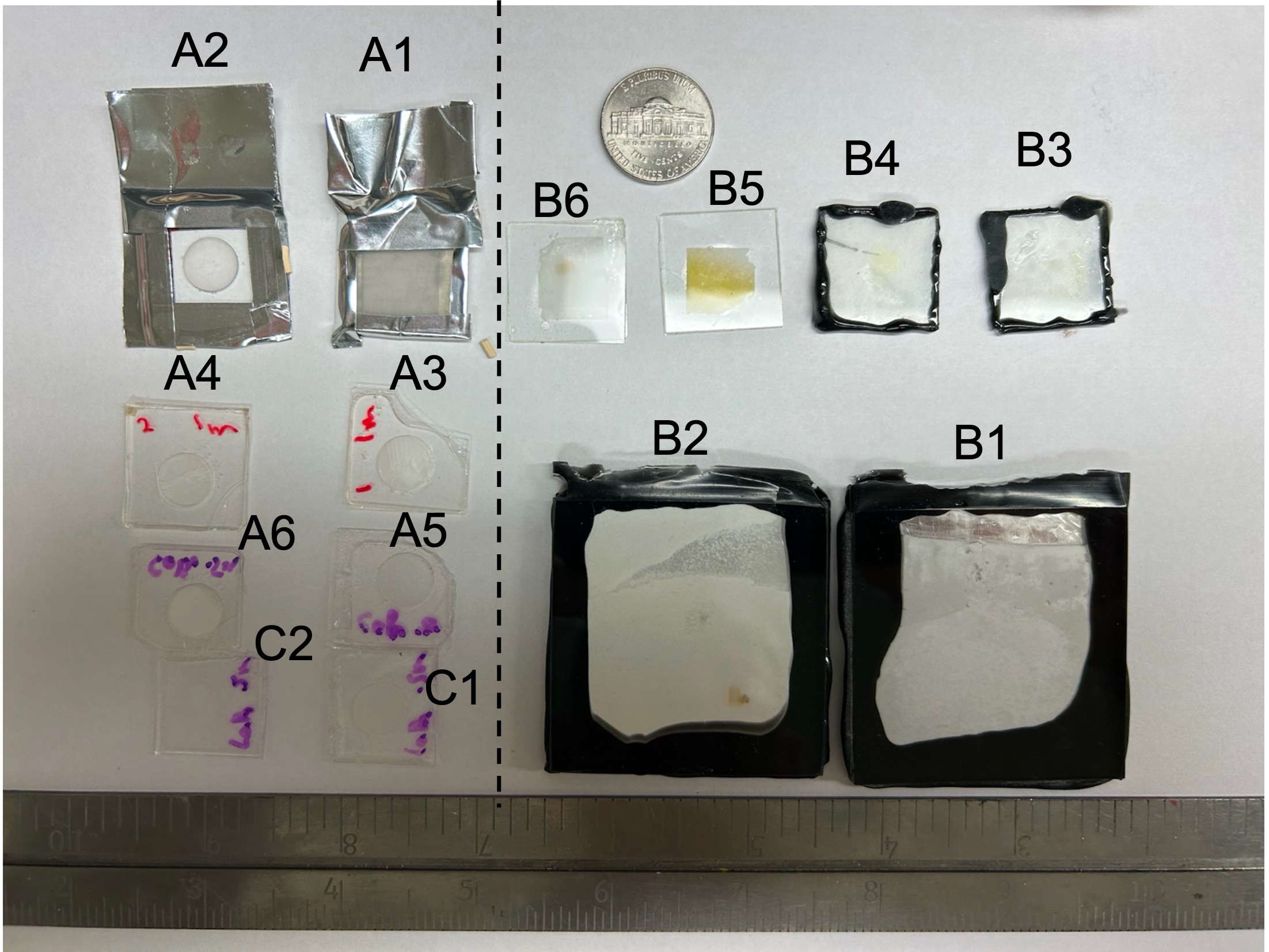}
\caption{Examples of prototype scintillator converters for X-ray phase contrast imaging using the 4th generation synchrotrons, such as APS-U at Argonne. The converters are based on two types of materials precursors, powder (labeled B1-B6, to the right of the vertical dashed line) and single crystals (labeled A1-A6, and C1-C2, to the left of the vertical dashed line). Single crystals are selected for high-resolution X-ray imaging. Additional details may be found in the main text, Sec.~\ref{sec:SC} and Sec.~\ref{sec:AHS}.}
\label{fig:SCINT_sam}
\end{figure}

Some examples of assembled scintillator converters (prototypes) for X-ray imaging are shown in Fig.~\ref{fig:SCINT_sam}. We may divide the prototypes into two top classes, based on the materials precursors used: powder converters and single-crystal convertors. The sub-classes in the powder converters may include the total converter area, powder size, the powder materials (LYSO, YAP, CeBr$_3$, LaBr$_3$:Ce, etc.),  the type of windows used (glass, quartz, and others), the window thicknesses, and the type of hermetic seal (epoxy compounds of different kinds.). The sub-classes in the single crystal class may include, the total converter area, the single crystal material (CeBr$_3$, LaBr$_3$:Ce), the crystal thicknesses, the window types and thicknesses, and the type of hermetic seal (epoxy compounds).

In Fig.~\ref{fig:SCINT_sam}, we have included 14 examples based on the above classification scheme. B1-B6 are `powder assemblies' (assemblies used powder scintillator precursors), A1-A6 (CeBr$_3$ crystals, various thicknesses) and C1-C2 (LaBr$_3$:Ce, 500 \textmu m thick) are crystal converters. B1 (CeBr$_3$ powder) and B2 (YAP:Ce powder) were the earliest prototypes, which had relatively large sizes ($\sim$ 10 cm long and wide) and used glass windows. One intension of such a large assemblies was for duel use: proton radiography, cm-size proton beam at 800 MeV~\cite{MACC:2024} and ultrafast X-ray imaging, mm-size partially coherent X-ray beam. However, since the 800-MeV proton attenuation and 20-30 keV X-ray attenuation, the spatial resolution requirement ($> 100$ \textmu m for the protons, and $<$ 10 \textmu m for the X-rays), and the flux densities are significantly different. The X-ray detection has since been focusing on using single crystals after a few additional powder assembly attempts: B3 and B4 (CeBr$_3$ powder, $\sim$ 70 $\mu$m grain sizes. Devcon 10 Minute black epoxy \#14255 from McMaster Carr. We learnt that the black epoxy is unlikely to provide a good long term seal from water infusion), B5 and B6 (CeBr$_3$ powder, $\sim$ 35 $\mu$m grain sizes. Transparent epoxy, Epotek 301, which is UV transmissive, radiation hard, and good for keeping Water out). Epotek epoxy was dried under a rough vacuum for $\sim$24 hours and stored under dry nitrogen to prevent dissolved moisture from interacting with the hygroscopic scintillators. This method is highly effective at removing moisture. By contrast, Devcon epoxy was also used in the dry box, but its drying efficacy remains unclear. To dry the dispensing tubes, they were placed in the dry box for approximately one week.

We started with a relatively thick CeBr$_3$ crystal A1 (2 mm thick) and large size ($\sim$5 cm). The area of the crystal scintillator has since reduced to 1 cm, which is sufficient for DCS/APS-U experiment for now (to minimize the edge emission effects of the scintillators, a larger area may be better). The thinnest scintillators are currently at 500 \textmu m for both CeBr$_3$, A5 and A6, and LaBr$_3$, C1 and C2. Prototypes of intermediate thicknesses (1 mm CeBr$_3$), A2, A3, A4, were also constructed.

\begin{figure}[!bht]
\centering
\includegraphics[width=3.6 in]{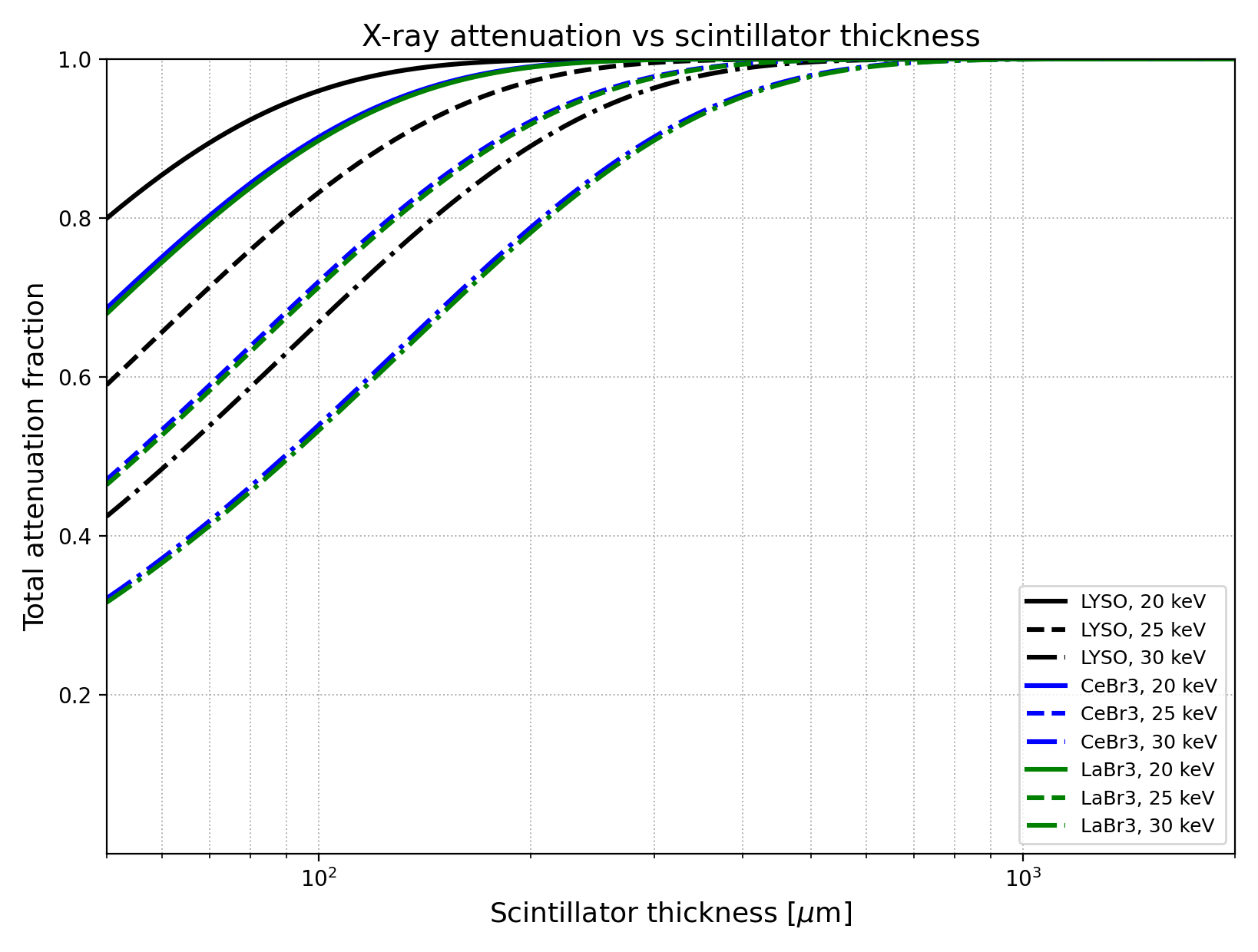}
\caption{Theoretical X-ray attenuation as a function of scintillator materials type (LYSO, LaBr$_3$:Ce, and CeBr$_3$), thickness (50 to 1000 \textmu m) and X-ray energies (20, 25 and 30 keV). Thicknesses in the range of 200 - 500 \textmu m scintillator crystal are sufficient for efficient absorption of X-rays and imaging at APS-U.}
\label{fig:SCINT_at1}
\end{figure}

The optimal scintillator thickness is to strike a balance between the X-ray attenuation length, Fig.~\ref{fig:SCINT_at1} and Table.~\ref{tab:properties}, and the imaging spatial resolution. Thicker scintillators have higher attenuation length. More than 99\% X-ray attenuation can be obtained for thickness $>$ 700 \textmu m and X-ray energies less than 30 keV. Thinner scintillators are preferred for higher spatial imaging resolution. The optimal thickness is therefore 2-3 times the 1/e-attenuation length, which corresponds to $\sim$140 to 210 \textmu m for both CeBr$_3$ and LaBr$_3$:Ce at 25 keV of X-ray energy.

\subsection{Light-yield characterization \label{sec:ly}}
The generally accepted light yields of the scintillators, summarized in Table~\ref{tab:properties}, can be affected by several factors in assemblies based on hygroscopic scintillators. These may include the optical windows, whose refractive index, thickness, internal reflections at interfaces, UV absorption, and scattering can influence light transmission, as well as additional attenuation and scattering along the optical path between the scintillator and the imaging detector, for example in the focusing optics. Therefore, we performed a series of light-yield measurements to characterize the actual light yield achieved with different assemblies.

\begin{figure}[!ht]
\centering
\includegraphics[width=3.6 in]{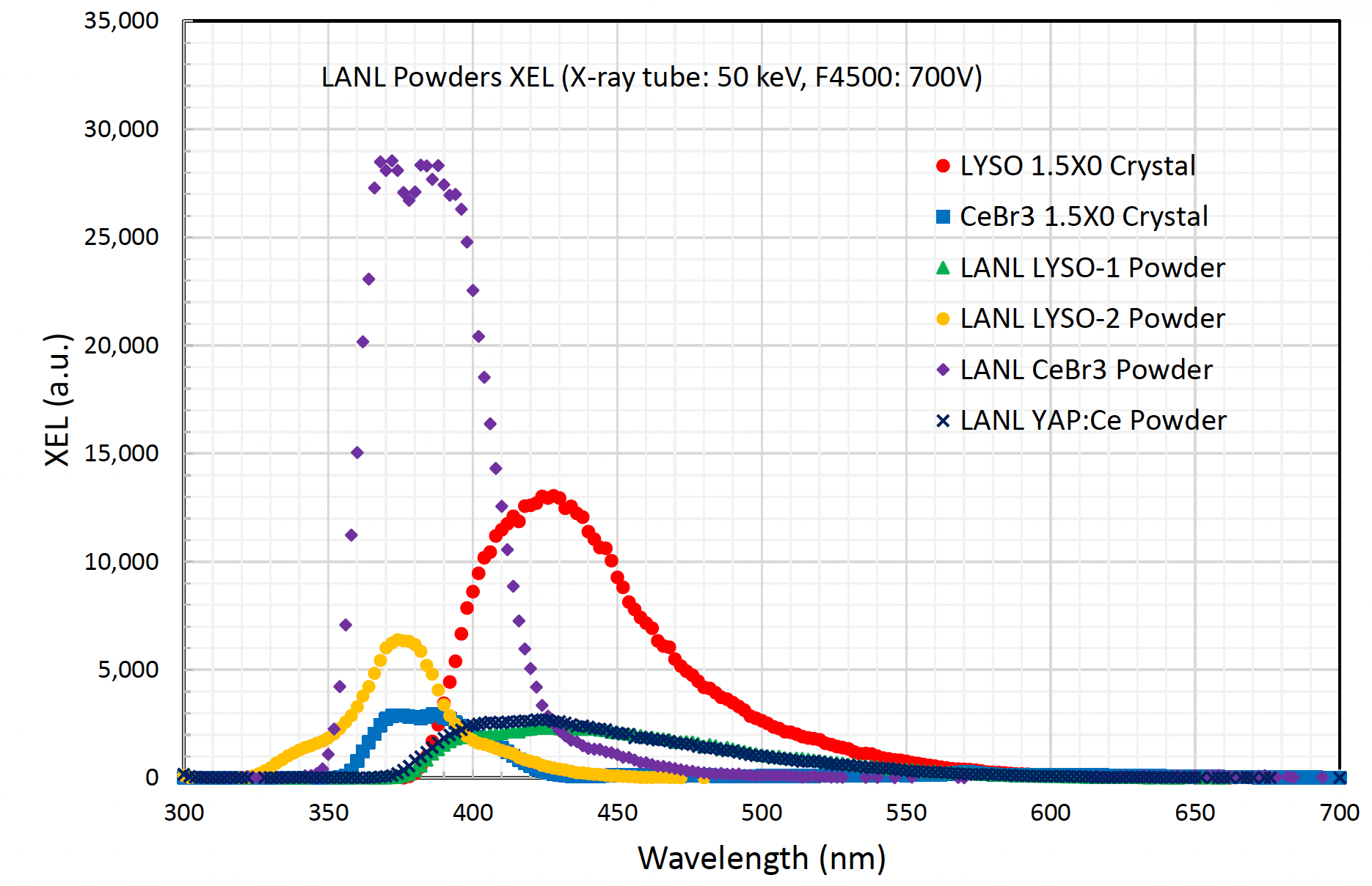}
\caption{Light yield variation as a function of emitted scintillator light wavelength for different powder samples, in comparison to a LYSO crystal scintillator (1.5 cm thick) and a partially hygroscopically degraded CeBr$_3$ crystal.}
\label{fig:Caltech}
\end{figure}

Light yield as a function of the emission wavelength was first measured using a Perkin Elmer Lambda-950 spectrometer (wavelength range: 175 nm to 3300 nm)
equipped with double-beam, double-monochromator, a Photomultiplier Tube (PMT) detector for the UV/Vis range, and a
general purpose optical bench~\cite{HZZC:2018}. The results are summarized in Fig.~\ref{fig:Caltech}. An Amptek Mini-X2 X-Ray Tube  (70 kV, 10 W) was used to excite the samples. The measurements confirmed that the peak emission wavelengths of CeBr$_3$ and LYSO:Ce as listed in Table.~\ref{tab:properties}. The two brightest emitters are the CeBr$_3$ powder assembly (B1 in Fig.~\ref{fig:SCINT_sam}) and a LYSO crystal sample (without the windows). YAP:Ce powder did not have enough emission intensity to warrant additional work. The emission from a CeBr$_3$ crystal (1.5 cm thick) is significantly less, in part due to hygroscopic degradation over time (in storage for more than a year). 

Spectral-integrated light-yield were then measured using a PMT detector (ET Enterprises 9111B series with a plano-concave window, a blue-green sensitive bialkali photocathode) without the Perkin Elmer Lambda-950 spectrometer and a variable energy X-ray source (Amersham model AMC 2084/2004) that emits the characteristic $k_\alpha$ and $k_\beta$ energies of six elements, Cu, Rb, Mo, Ag, Ba and Tb. The Amersham source was used in an earlier development of a billion pixel X-ray camera~\cite{WABD:2021}. Individual pulse shape analysis from the PMT confirmed the characteristic decay time as shown in Table.~\ref{tab:properties}.

\begin{figure}[!t]
\centering
\includegraphics[width=3.6 in]{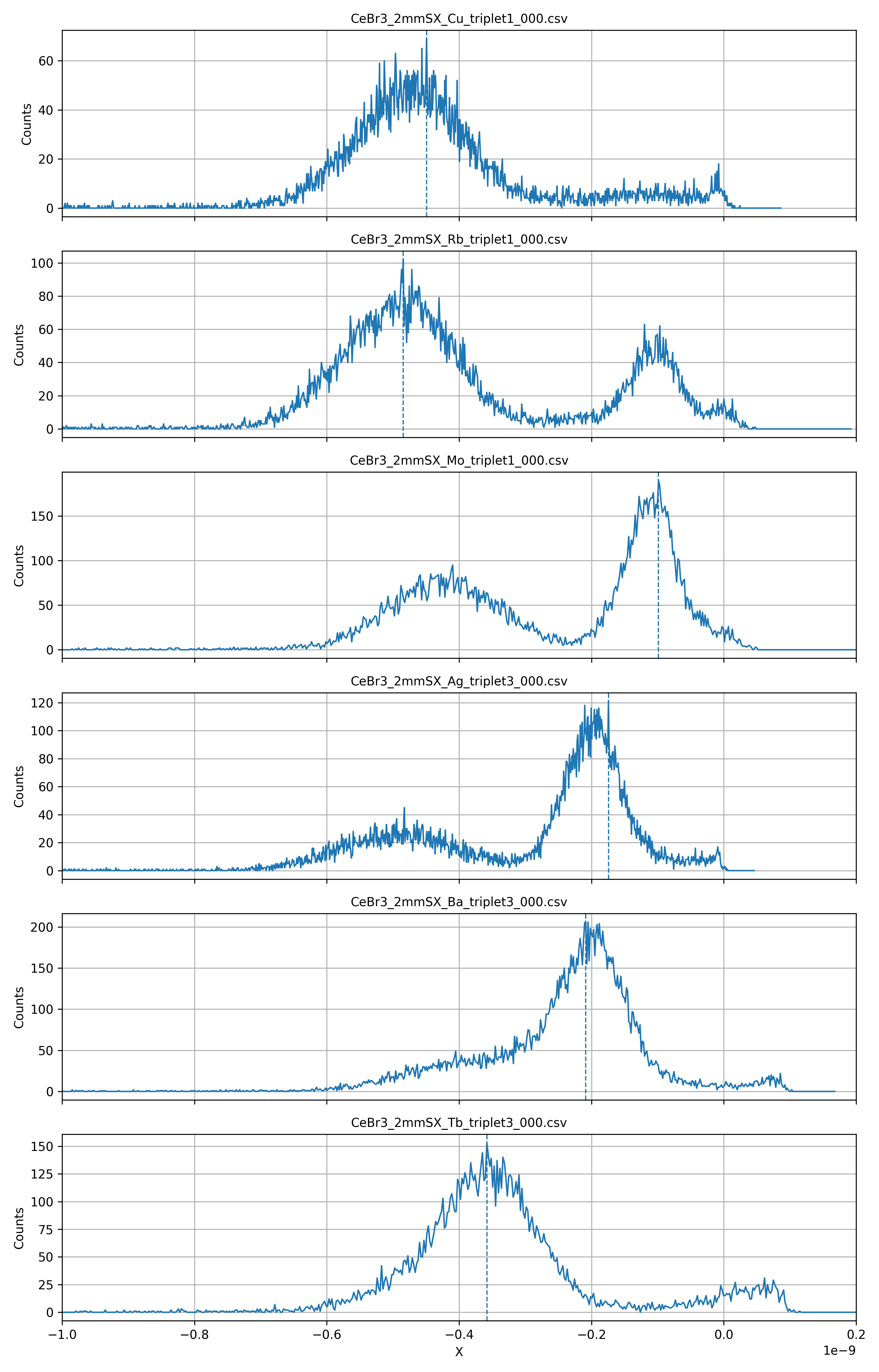}
\caption{CeBr$_3$ cystal (Assembly A1 in Fig.~\ref{fig:SCINT_sam}, 10-mm diameter, 2-mm nominal thickness, vendor data) light yield as a function of $k_\alpha$ and $k_\beta$ characteristic X-ray of various elements: Cu, Rb, Mo, Ag, Ba, and Tb (from top to bottom). The light yield (pulse area, horizontal axis, in mV$\cdot$s) is confirmed to be linearly proportional to individual X-ray photon energies.}
\label{fig:CeBr3_2mma}
\end{figure}

\begin{figure}[!t]
\centering
\includegraphics[width=3.6 in]{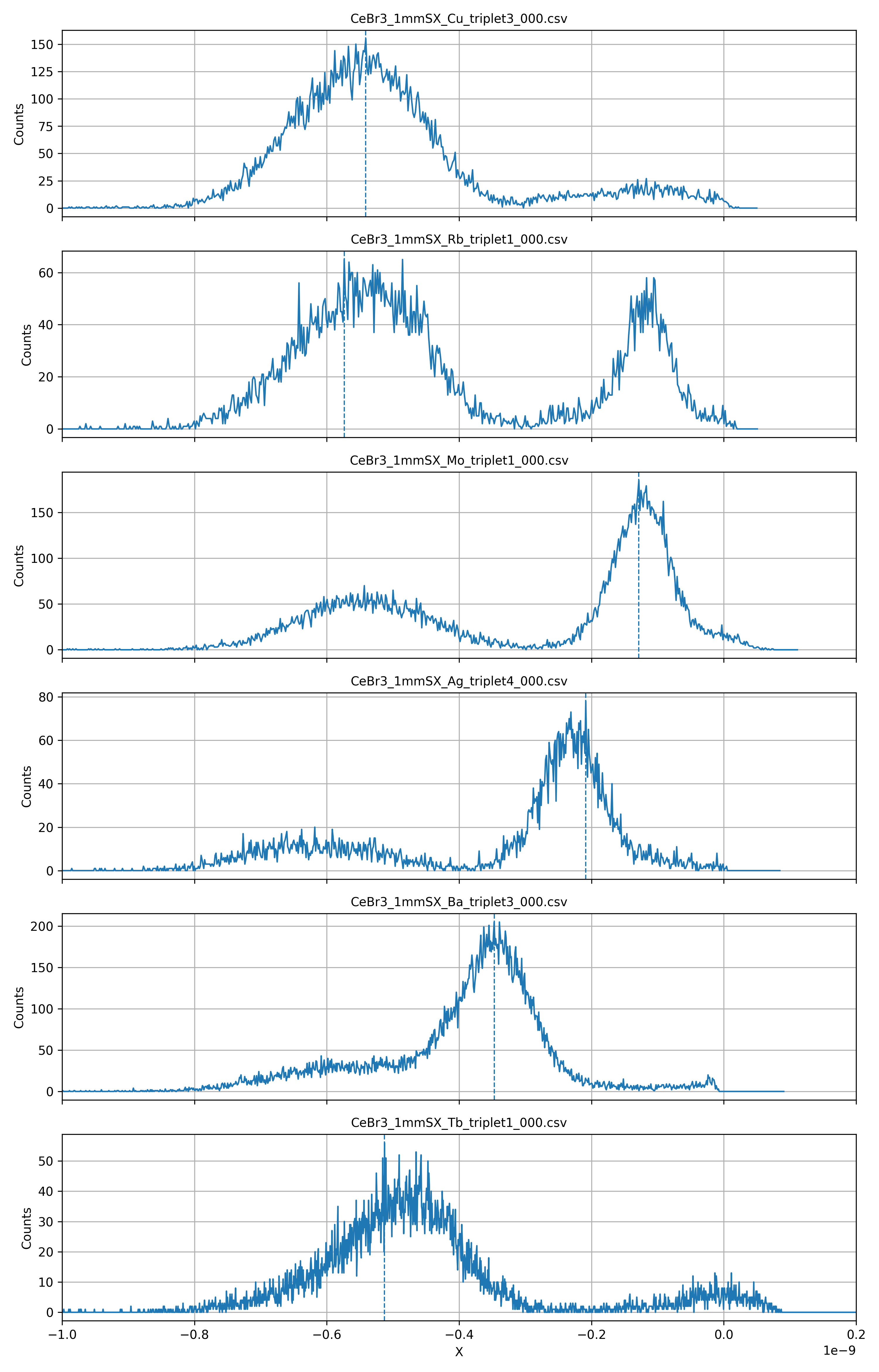}
\caption{CeBr$_3$ cystal (Assembly A2 in Fig.~\ref{fig:SCINT_sam}, 10-mm diameter, 1-mm nominal thickness, vendor data) light yield as a function of $k_\alpha$ and $k_\beta$ characteristic X-ray of various elements: Cu, Rb, Mo, Ag, Ba, and Tb (from top to bottom). The light yield (pulse area, horizontal axis, in mV$\cdot$s) is confirmed to be linearly proportional to individual X-ray photon energies. The total light detected is greater than the thicker scintillator as shown in Fig.~\ref{fig:CeBr3_2mma}.}
\label{fig:CeBr3_1mma}
\end{figure}

Histograms of the total scintillator light yield induced by different characteristic X-ray energies, as measured by the pulse height $V(t)$ integrated over time, $\int V(t) dt$, are summarized in Figs.~\ref{fig:CeBr3_2mma} and \ref{fig:CeBr3_1mma} for two CeBr$_3$ thicknesses (2 mm and 1 mm, respectively).  The negative values of the integrals (horizontal values in Figs.~\ref{fig:CeBr3_2mma} and \ref{fig:CeBr3_1mma}) are due to the negative going pulses from the PMT detector. 

We highlight two observations based on the data. 1.) Independent of X-ray energies, we see that overall, the total light collected by the detector increases when the CeBr$_3$ scintillator thickness decrease from 2 mm to 1 mm. 2.) The average total light yield is proportional to the X-ray input energy, for both thicknesses. The proportionality is, nevertheless, dependent on the thickness, indicating potentially thickness-dependent light loss mechanisms. Additional studies on thinner CeBr$_3$ and LaBr$_3$:Ce at 0.5 mm thickness (A5, A6, and C1, C2 in Fig.~\ref{fig:SCINT_sam}) will be published elsewhere.

\section{Experimental results in X-ray imaging}
We now highlight the results of X-ray imaging using the prototype scintillator assemblies described above.  For static imaging, Sec.~\ref{sec:static}, we used the Amersham variable energy X-ray source. For dynamic X-ray phase contrast imaging, Sec.~\ref{sec:Dynamic}, we performed the experiments at the Dynamic Compression Sector (DCS) in the Advanced Photon Source Upgrade (APS-U) facility of Argonne National Laboratory, leveraging the gas-gun-driven projectiles at impact velocities ranging from a few hundred m/s to a few km/s. %Describe the detector, electronics, simulation, fabrication, irradiation setup, data acquisition, calibration, and analysis methods as appropriate for your paper.

\begin{figure}[!t]
\centering
\includegraphics[width=3.6 in]{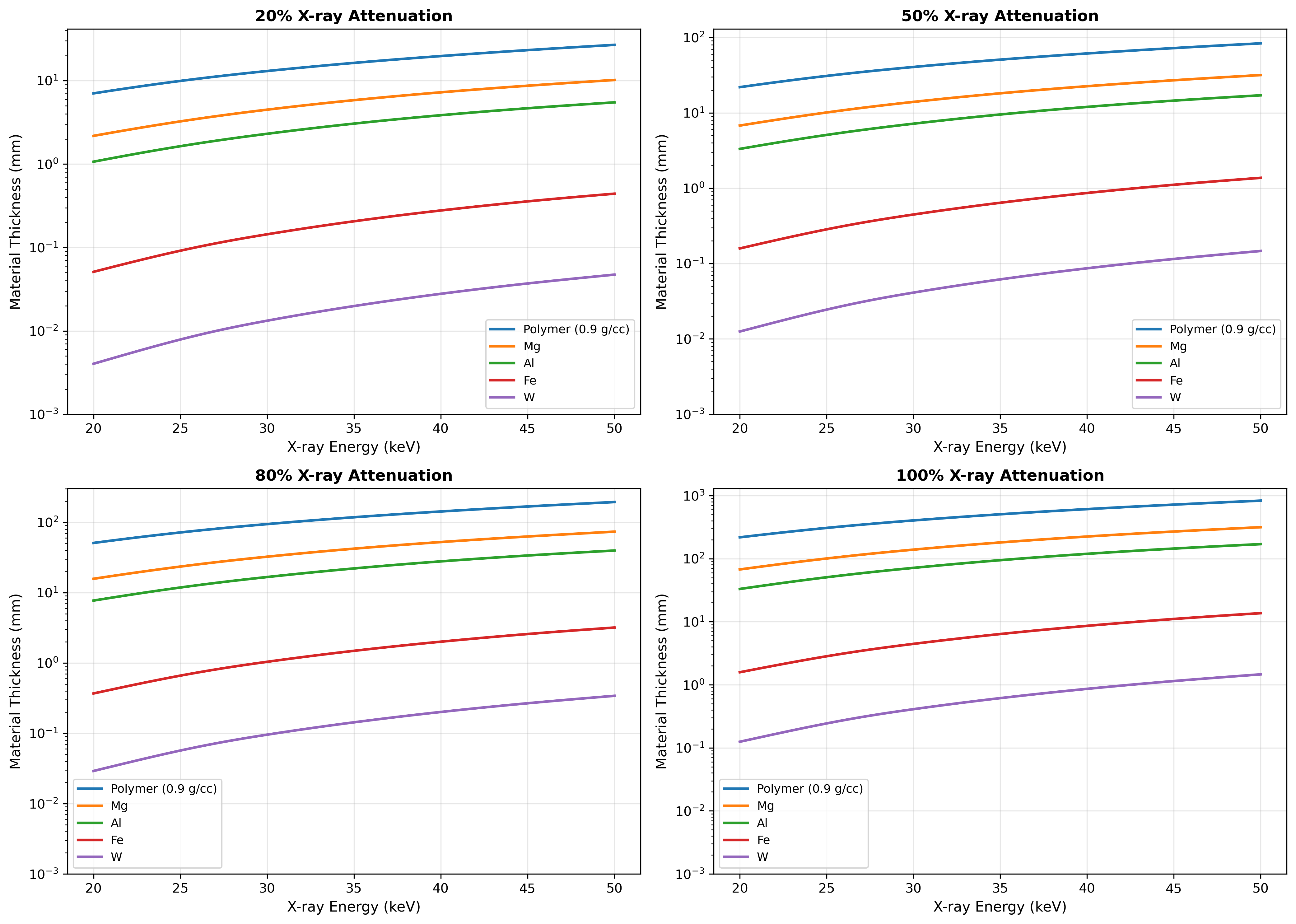}
\caption{X-ray attenuation curves as a function of X-ray energy, in the range of 20-50 keV, and for different materials, polymer at 0.9 g/cm$^3$ density, Mg, Al, Fe and W. The metals at their normal solid densities. For X-ray PCI, the object attenuation in the range of $\sim$ 30-70\% would determine the good object sample size along the X-ray beam.}
\label{fig:Attenuation_1}
\end{figure}

The experiments are guided by theoretical X-ray attenuation analysis as illustrated in Fig.~\ref{fig:Attenuation_1}. In the X-ray energy range from 20 to 50 keV, which matches the energies range expected for imaging at APS-U, the thicknesses of the materials that can be X-ray radiographed strongly depend on the atomic number $Z$ (or the mean $Z$ for a compound) and density. For 50\% attenuation, for example, W thicknesses are limited to about 10 \textmu m at 20 keV X-ray energy to about 100 \textmu m at 50 keV. Polymers, assuming a density of 0.9 g/cm$^3$, can be more than 1000 times thicker at 50\% attenuation, ranging from 2 cm at 20 keV to about 10 cm at 50 keV.
%\subsection{Example Figure}

\subsection{Static imaging with a laboratory X-ray source \label{sec:static}}
\begin{figure}[!t]
\centering
%\fbox{\rule{0pt}{1.6in}\rule{0.95\linewidth}{0pt}}
\includegraphics[width=3.6 in]{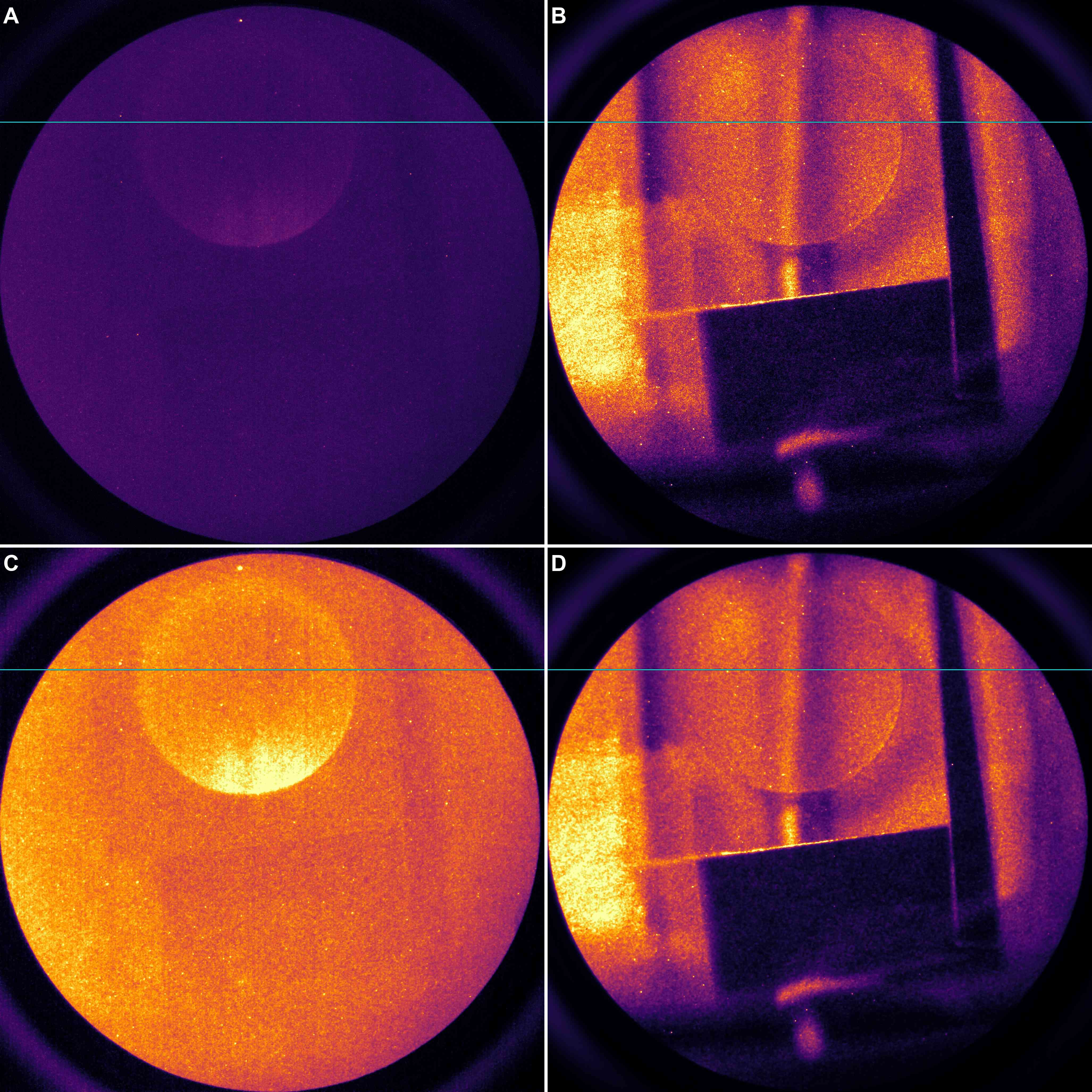}
\caption{A. A static X-ray induced image of a wire assembly using a 1-mm thick CeBr$_3$ crystal, sample A2 in Fig.~\ref{fig:SCINT_sam}. The X-ray source is at Ag characteristic energies $k_\alpha$ and $k_\beta$. The time of integration was 1 s. B. The corresponding visible image of the same setup. The time of integration was 1 \textmu s. C and D are processed images corresponding to A and B respectively. All raw images are in black-and-white intensity maps, and shown here in false color. The image contrast are limited by the total X-ray flux of the Amersham source.}
\label{fig:Ag_Vis_CeBr3}
\end{figure}

Static imaging results of a wire assembly using the CeBr$_3$, the 1-mm thick sample A2 in Fig.~\ref{fig:SCINT_sam}, is shown in Fig.~\ref{fig:Ag_Vis_CeBr3}. A PCO C1 intensified camera with a maximum gain at 800 V was used in taking both the X-ray and visible light images. Additional denoising and smoothing of the raw images (in false color), Fig.\ref{fig:Ag_Vis_CeBr3}A and Fig.\ref{fig:Ag_Vis_CeBr3}B, are shown as Fig.\ref{fig:Ag_Vis_CeBr3}C and Fig.\ref{fig:Ag_Vis_CeBr3}D.

\begin{figure}[!t]
\centering
\includegraphics[width=3.6 in]{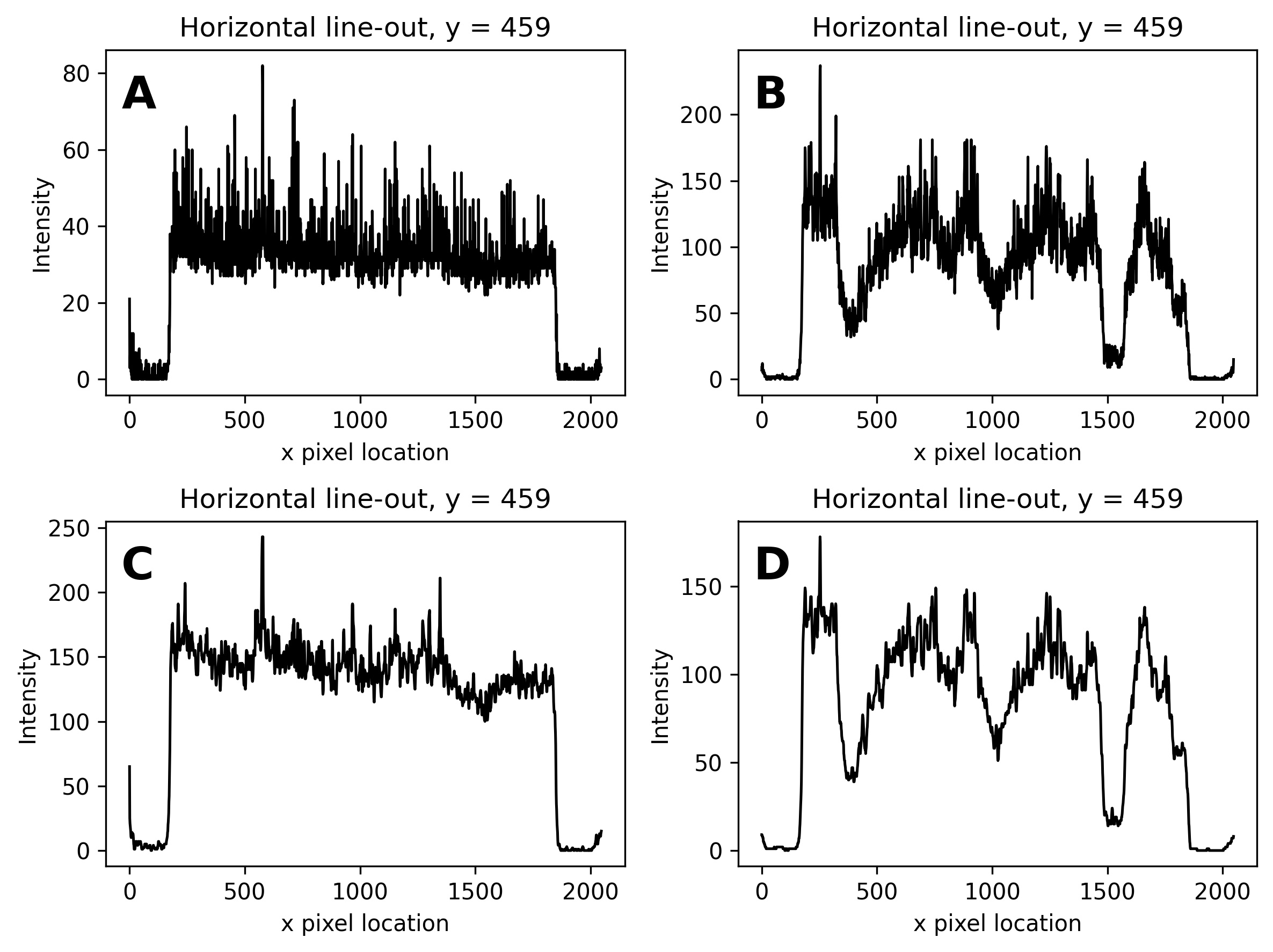}
\caption{Lineout analysis for the image pairs in Fig.~\ref{fig:Ag_Vis_CeBr3}. The horizontal lineout location is marked in Fig.~\ref{fig:Ag_Vis_CeBr3}. The image contrast are limited by the total X-ray flux of the Amersham source.}
\label{fig:Ag_Vis_CeBr3_lo1}
\end{figure}

Despite of the integration time of 1 s, the signal-to-noise ratio (S/N) of the features are relatively weak ($\sim <$ 1) for the X-ray images. This is better illustrated in the horizontal lineout plot, Fig.~\ref{fig:Ag_Vis_CeBr3_lo1}. The location of the lineout was slightly above the center of the scintillator circle. Further improvements of the S/N will be prioritized for a.) constructing thinner CeBr$_3$ assemblies (200 to 500 $\mu$s), and b.) using a higher intensity X-ray source such as the APS-U synchrotron.

\begin{figure}[!t]
\centering
\includegraphics[width=3.6 in]{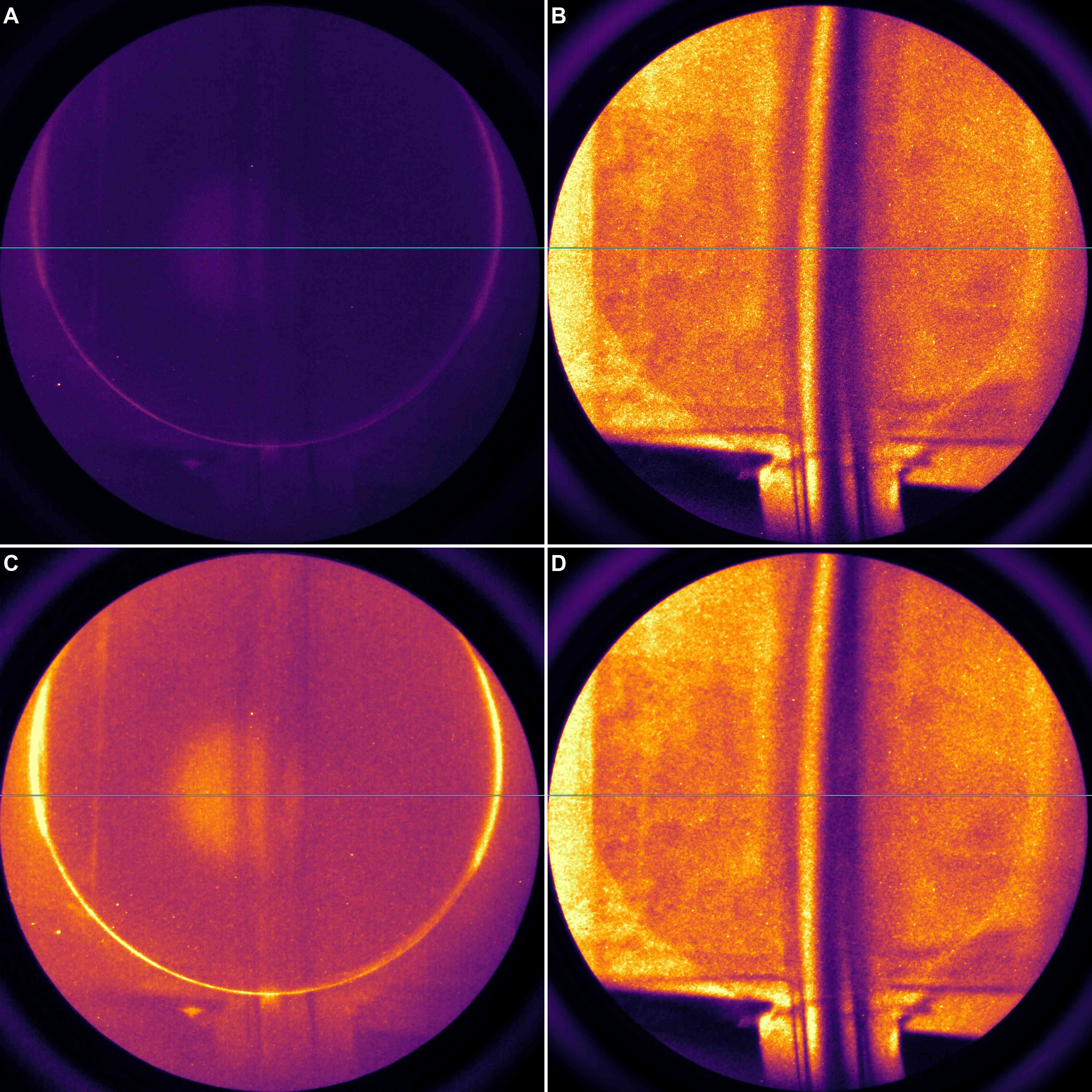}
\caption{Laboratory X-ray Imaging using Ag $k_\alpha$ and $k_\beta$ emission as the source and a 200-\textmu m thick LYSO scintillator as the X-ray-to-visible converter. The image contrast, also limited by the total X-ray flux of the Amersham source, is improved over the thicker CeBr$_3$ as shown in Fig.~\ref{fig:Ag_Vis_CeBr3}.}
\label{fig:Ag_Vis_LYSO}
\end{figure}

\begin{figure}[!t]
\centering
\includegraphics[width=3.6 in]{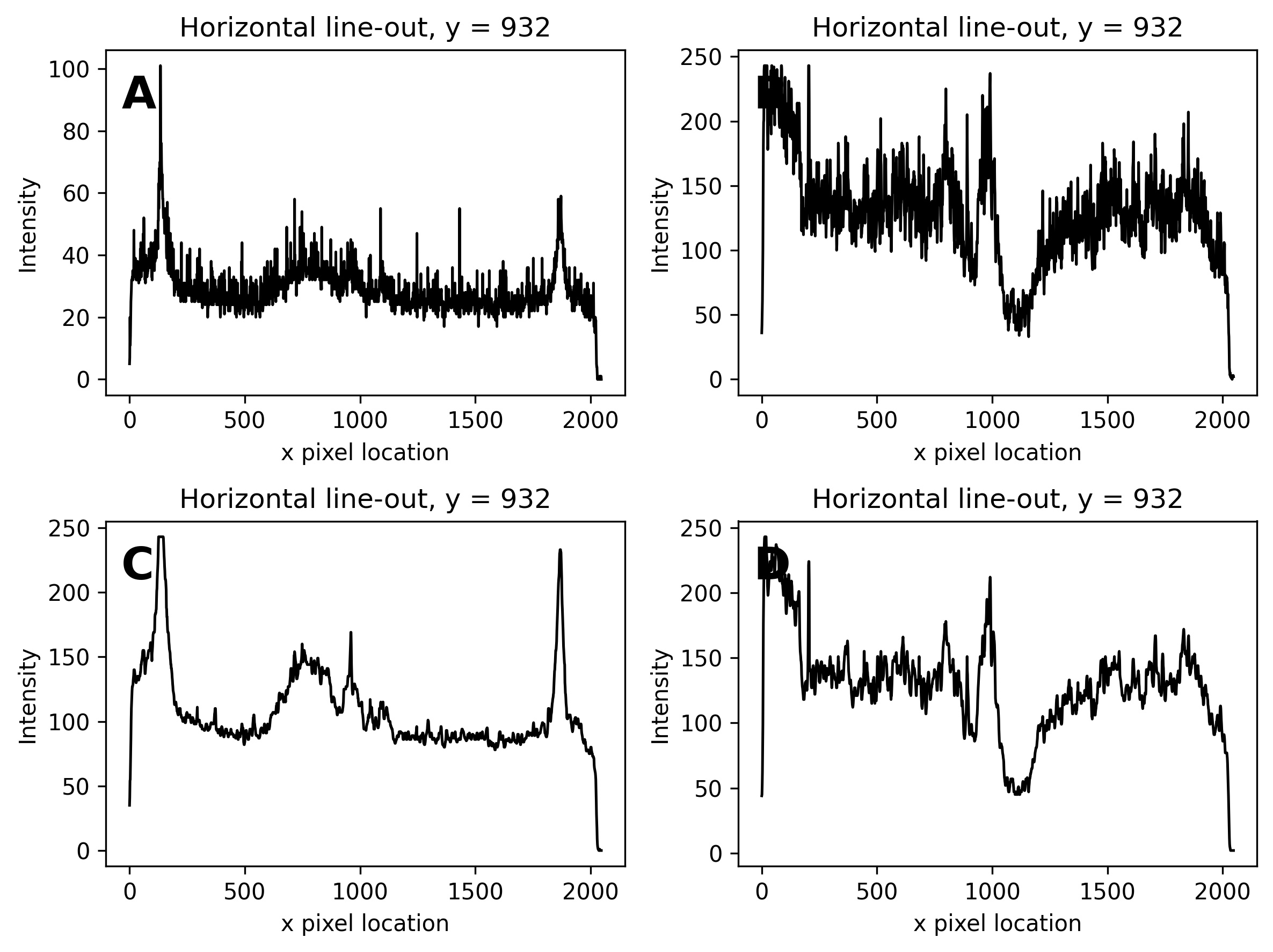}
\caption{Lineout analysis for the image pairs in Fig.~\ref{fig:Ag_Vis_LYSO}. The image lineout contrast, also limited by the total X-ray flux of the Amersham source, is improved over the thicker CeBr$_3$ as shown in Fig.~\ref{fig:Ag_Vis_CeBr3_lo1}.}
\label{fig:Ag_Vis_LYSO_lo1}
\end{figure}

In comparison to CeBr$_3$ results, Fig.~\ref{fig:Ag_Vis_LYSO} shows similar static imaging results using a 200-\textmu m thick LYSO scintillator and the same X-ray source and the same wire array as the static target of observation. The corresponding lineout plots are shown in Fig.~\ref{fig:Ag_Vis_LYSO_lo1}.

\subsection{Dynamic X-ray imaging using APS-U \label{sec:Dynamic}}
\begin{figure}[!t]
\centering
\includegraphics[width=3.6 in]{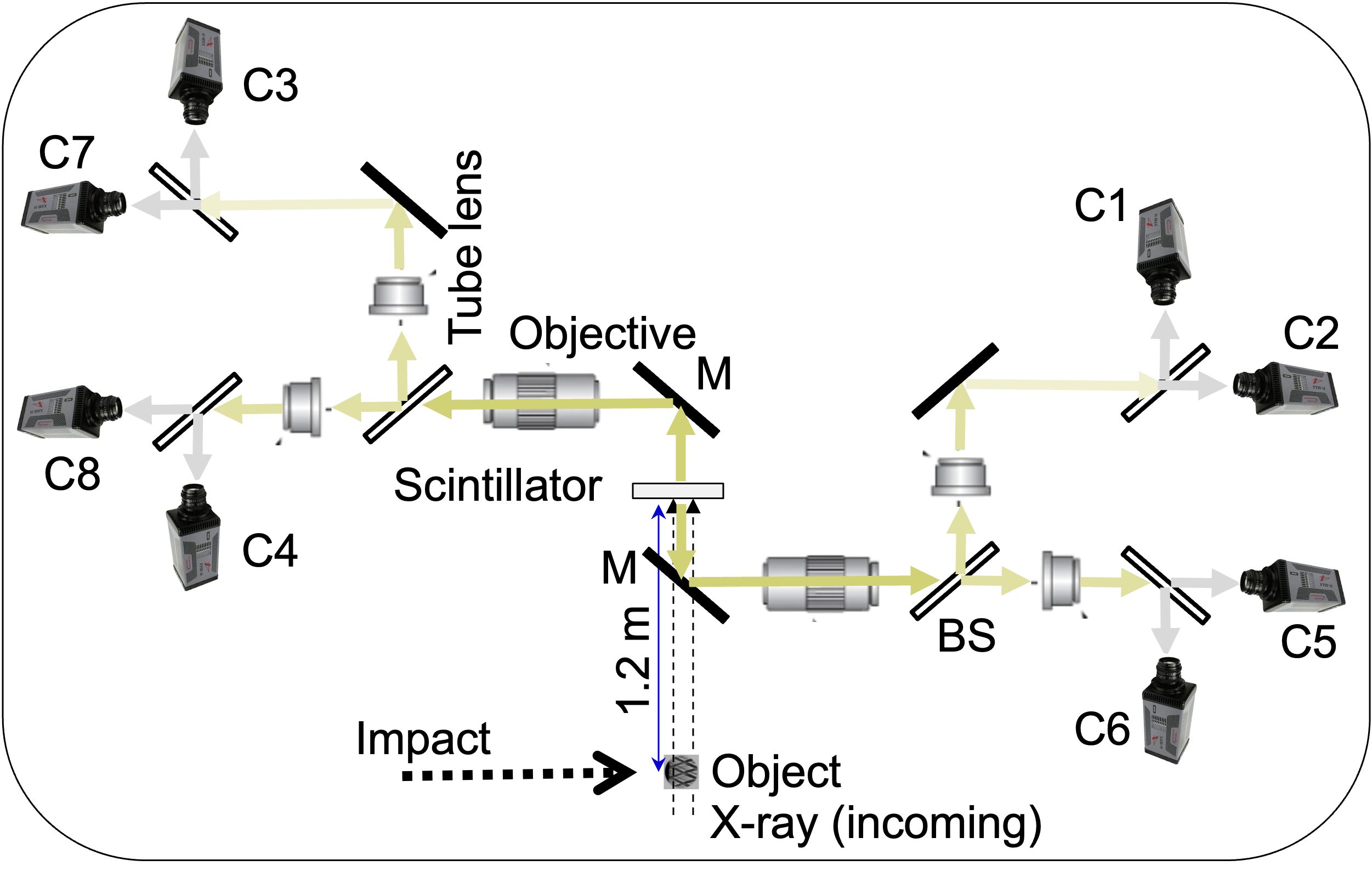}
\caption{APS/DCS 8-camera setup for dynamic X-ray Phase Contrast Imaging (PCI). The camera model is Princeton Instrument PI-MAX4. The microscope optics (UV enhanced) gives a factor of $\times$10 magnification of the features at the scintillator location. Half of the camera collects scintillator light from the front side (relative to the incoming X-ray flux) of the scintillator, C1, C2, C5 and C6. The other half (C3, C4, C7 and C8)  from the back side.}
\label{fig:APS1}
\end{figure}

\begin{figure}[!t]
\centering
\includegraphics[width=3.6 in]{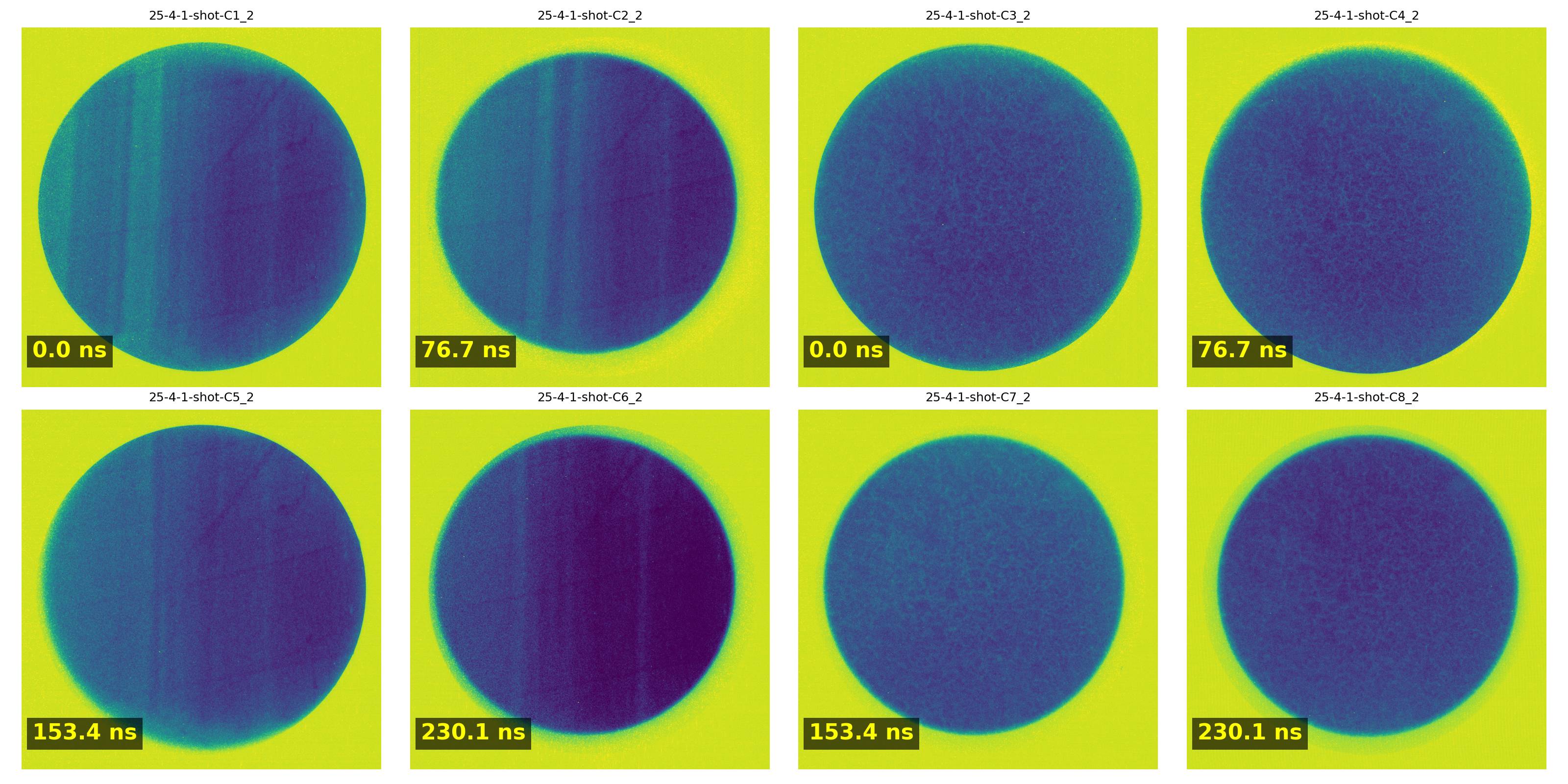}
\caption{Dynamic propagation of a density gradient observed at a frame rate of 13 MHz. The CeBr$_3$ scintillator crystal was 2-mm thick, similar to A1 in Fig.~\ref{fig:SCINT_sam}. The four frames of the images on the left are from the front side of the scintillator. The others are from the back side. One pair of cameras from each side of the scintillator are time-synced.}
\label{fig:APS2}
\end{figure}

The 8-camera configuration at DCS is shown in Fig.~\ref{fig:APS1}. Each camera can take up to 2 dynamic images, depending on the relative timing delays. For our initial test, each camera took only one images with relative delays, and generated a set of 8-frame `movie' as shown in Fig.~\ref{fig:APS2}.

\begin{figure}[!t]
\centering
\includegraphics[width=3.6 in]{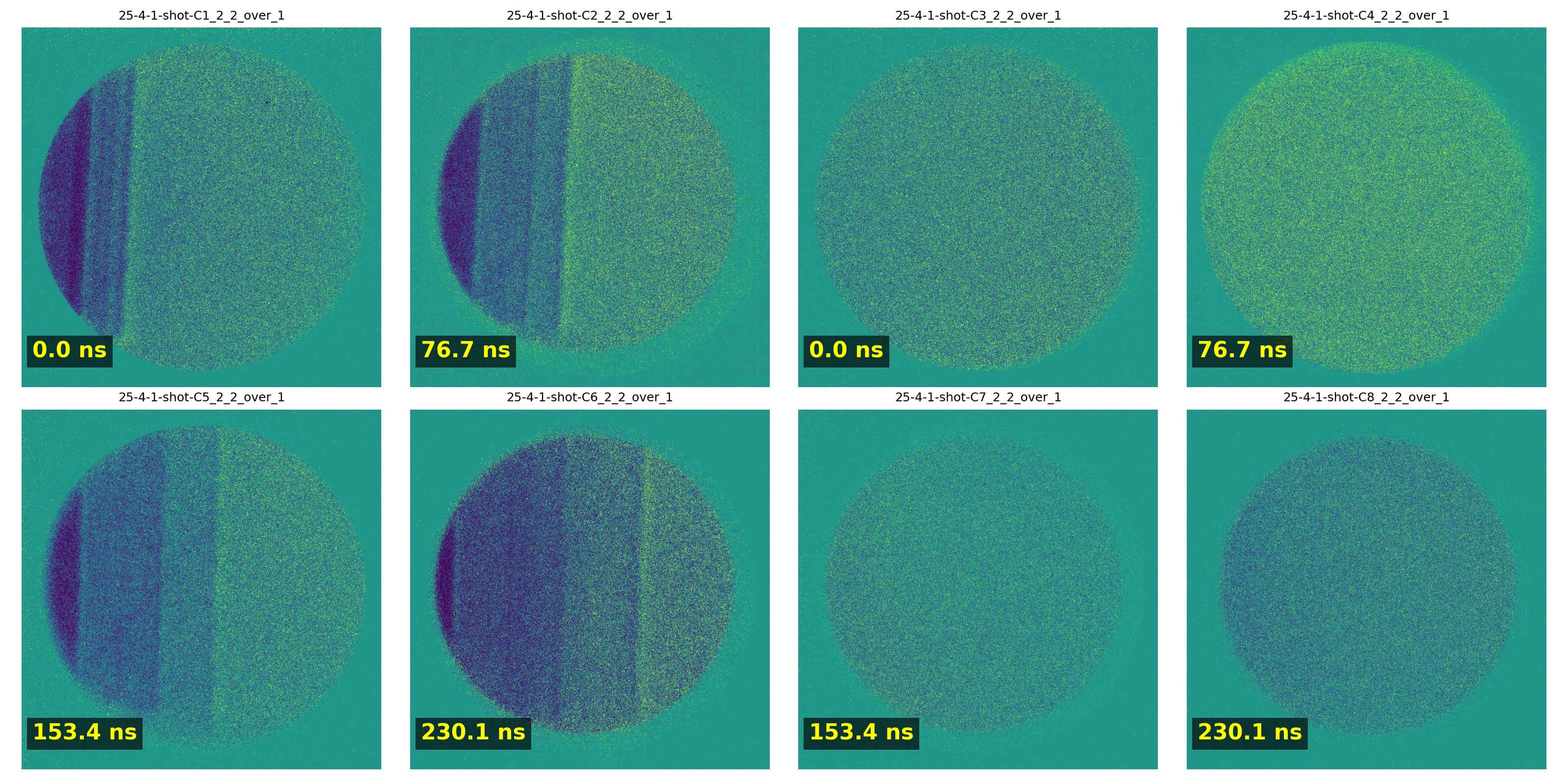} %{Cleaned25_4_049.png}
\caption{The signal-to-noise ratio (S/N) of the density gradient dynamics, as shown in Fig.~\ref{fig:APS2},  improved through background reduction techniques.}
\label{fig:50cleaned}
\end{figure}

Only an earlier version of the 2-mm thick CeBr$_3$ crystal (prior to the construction of A1 in Fig.~\ref{fig:SCINT_sam}) was available at the time of the DCS experiment. The results from one dynamic experiment using the 8 cameras are summarized in Fig.~\ref{fig:APS2}. Several observations can be made. A.) The scintillator are bright and fast enough for taking movies using continuous X-ray pulses at 76.7 ns intervals ($\sim$ 13 MHz), this is illustrated by the four consecutive X-ray images on the left side of the dashed line in Fig.~\ref{fig:APS2}. B.) The scintillator ($\sim$ 2 mm) was too thick, which washed out any interesting features from the back side of the scintillator. C.) Imaging processing techniques is useful to remove artifacts from defects of the scintillator and improve the S/N for the observed dynamic features, as shown in Fig.~\ref{fig:50cleaned}.

We further call out the density gradient propagation under the impact as the main dynamic feature observed in Fig.~\ref{fig:APS2} and Fig.~\ref{fig:50cleaned}. Further analysis of the density gradient propagation will be discussed elsewhere.

\section{Discussion}
The present results highlight both the promise and the practical challenges of using fast bromide scintillators for ultrafast hard X-ray imaging at fourth-generation synchrotron sources. LaBr$_3$:Ce and CeBr$_3$ are attractive because their scintillation decay times are substantially shorter than those of LYSO/LSO, while their light yields remain high enough for photon-starved MHz-rate measurements. These properties directly address one of the central requirements for APS-U dynamic materials experiments: minimizing inter-frame persistence while preserving sufficient signal-to-noise ratio at frame spacings on the order of 77 ns or below.

A key observation from this work is that noticeable hygroscopic degradation in CeBr$_3$ crystals is possible when the material is not adequately protected from ambient moisture. Such degradation can manifest as surface clouding, reduced optical transmission, degraded light extraction, and ultimately reduced image intensity and uniformity. This behavior is consistent with the known hygroscopic nature of halide scintillators and is an important practical distinction between bromide scintillators and oxide scintillators such as LYSO. However, the prototype packaging and handling studies presented here also indicate that this limitation can be mitigated. With appropriate encapsulation, moisture barriers, and optical coupling, CeBr$_3$ and LaBr$_3$:Ce can be made sufficiently stable for beamline testing and potential detector deployment.

The relevance of this result is significant for nuclear science instrumentation. In many laboratory nuclear spectroscopy applications, LaBr$_3$:Ce and CeBr$_3$ are used as bulk, hermetically sealed crystals coupled directly to photomultiplier tubes or silicon photomultipliers. In contrast, synchrotron imaging applications often require thin scintillator screens, large optical apertures, and free-space or lens-coupled readout geometries. These requirements expose the scintillator and its optical interfaces to different environmental and mechanical constraints. Therefore, demonstrating packaging approaches compatible with indirect X-ray imaging is an important step toward translating the excellent intrinsic properties of bromide scintillators into practical ultrafast beamline detectors.

The choice between CeBr$_3$ and LaBr$_3$:Ce involves several tradeoffs. LaBr$_3$:Ce is well known for its very high light yield and excellent energy resolution in gamma-ray spectroscopy, while CeBr$_3$ offers comparable fast timing behavior and avoids the intrinsic radioactivity associated with naturally occurring $^{138}$La. For ultrafast synchrotron imaging, intrinsic radioactivity is generally less important than in low-background nuclear spectroscopy because the X-ray signal during a pulse is large. However, background emission, dark-field contributions, and camera integration conditions may still need to be considered for certain low-signal or long-exposure measurements. CeBr$_3$ may therefore be favorable in applications where intrinsic background must be minimized, whereas LaBr$_3$:Ce remains attractive when maximum light yield is the primary requirement.

For MHz-rate hard X-ray movies, scintillator thickness is also a critical design parameter. Thicker CeBr$_3$ or LaBr$_3$:Ce screens increase X-ray absorption and light output, which can improve signal-to-noise ratio. However, increased thickness can also degrade spatial resolution because of lateral light spread, depth-of-interaction effects, and optical scattering. In addition, for very high frame-rate measurements, any slow scintillation component or delayed optical transport can contribute to image persistence. Therefore, thinner CeBr$_3$ or LaBr$_3$:Ce scintillators may be advantageous when spatial resolution and temporal fidelity are prioritized over absolute stopping power. The optimum thickness will depend on X-ray energy, magnification, numerical aperture, camera sensitivity, and the required balance between absorption efficiency and image sharpness.

\begin{figure}[!t]
\centering
%\fbox{\rule{0pt}{1.6in}\rule{0.95\linewidth}{0pt}}
\includegraphics[width=3.6 in]{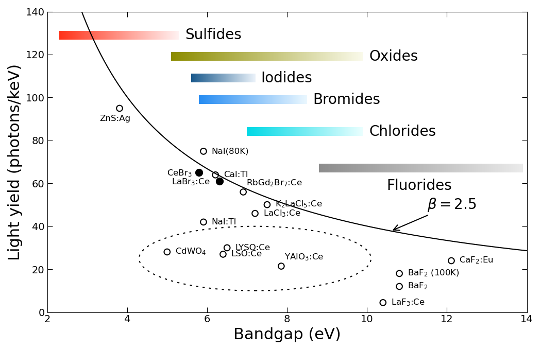}
\caption{Updated Dorenbos curve illustrating that CeBr$_3$ and LaBr$_3$:Ce still fit the similar relationship between scintillator material properties and light output for Ce$^{3+}$-activated scintillators~\cite{Dore:2002,PiLW:2019}.}
\label{fig:Doren1}
\end{figure}

The performance of LaBr$_3$:Ce and CeBr$_3$ can also be placed in the broader context of the well-known scintillator trends discussed by Dorenbos and subsequent studies~\cite{Dore:2002,PiLW:2019}, as summarized in Fig.~\ref{fig:Doren1}. The Dorenbos framework is useful because it relates scintillator performance to fundamental material parameters, including band-gap energy, activator emission energy, energy-transfer efficiency, and nonproportionality. In this context, Ce$^{3+}$-activated scintillators occupy an important region of parameter space: the allowed 5$d$--4$f$ transition of Ce$^{3+}$ provides fast emission, while host materials with high density and efficient carrier transport can produce large light output. LaBr$_3$:Ce and CeBr$_3$ are therefore not merely empirically attractive materials; their performance is consistent with established trends for high-yield, fast Ce$^{3+}$ scintillators.

The comparison with previous work also clarifies the role of the detector system as a whole. In gamma-ray spectroscopy, the excellent energy resolution of LaBr$_3$:Ce and CeBr$_3$ is often emphasized. For ultrafast synchrotron imaging, however, the most important metrics are not only light yield and decay time, but also optical extraction efficiency, emission wavelength compatibility with the camera, radiation tolerance, screen uniformity, and packaging robustness. The emission of these bromide scintillators is concentrated in the near-UV/blue spectral range, where the quantum efficiency of the detector and the transmission of optical elements can vary substantially. Thus, the scintillator cannot be optimized independently from the lens, window, coupling medium, and sensor. The best performance will require integrated optimization of the scintillator--optics--camera chain.

Several limitations of the present study should be noted. First, the beamline demonstrations represent initial tests rather than a complete qualification of long-term performance under all APS-U operating conditions. Extended exposure studies are needed to quantify radiation damage, possible color-center formation, and changes in light yield or decay kinetics after repeated high-flux X-ray irradiation. Second, the present packaging approaches should be evaluated over longer storage and operation times, including thermal cycling, humidity exposure, and mechanical handling. Third, a systematic comparison of scintillator thickness, surface finish, reflector configuration, and optical coupling was not fully completed. These variables can strongly affect both light output and spatial resolution. Finally, the results obtained with a particular camera and optical configuration may not directly translate to other beamline instruments without further optimization.

Despite these limitations, the present work supports the use of CeBr$_3$ and LaBr$_3$:Ce as practical candidates for next-generation indirect detectors at APS-U and similar sources. Their combination of high density, high light output, and fast decay provides a compelling route to reducing motion blur and frame-to-frame contamination in ultrafast dynamic materials experiments. This is particularly important for irreversible phenomena such as shock compression, fracture, phase transformation, ejecta formation, and rapid mesoscale deformation, where repeated measurements under identical conditions may not be possible. In such experiments, the ability to record a sequence of high-quality images within a single event is often more valuable than optimizing any single-frame metric.

In addition to bromide single crystals, emerging scintillator materials may provide complementary capabilities. Metal-halide perovskites are of interest because they can exhibit high light yield, tunable emission wavelength, fast radiative recombination, and compatibility with thin-film or structured-screen fabrication. Their processability may be beneficial for producing large-area or application-specific scintillator geometries. However, many perovskites also face challenges related to environmental stability, ion migration, afterglow, toxicity, and radiation hardness. These issues must be addressed before they can be considered robust alternatives for high-flux synchrotron operation.

High-entropy scintillators and high-entropy oxide or halide thin films represent another promising direction. By incorporating multiple cations or anions into a single host lattice, high-entropy materials may enable tuning of band structure, defect populations, carrier localization, and radiation tolerance. Such materials could potentially combine fast emission with improved mechanical, thermal, or radiation stability. At present, however, their scintillation mechanisms, reproducibility, and scalability remain less mature than those of commercial LaBr$_3$:Ce and CeBr$_3$. For this reason, they are best viewed as longer-term candidates that may complement, rather than immediately replace, bromide scintillators.

Overall, the results suggest that the path toward MHz-rate scintillator-based imaging will require both material selection and detector engineering. LaBr$_3$:Ce and CeBr$_3$ provide favorable intrinsic scintillation properties, but their successful deployment depends on moisture-resistant packaging, efficient light collection, spectral matching to high-quantum-efficiency sensors, and careful selection of scintillator thickness for the intended X-ray energy and spatial-resolution requirement. Continued optimization along these directions will be essential for realizing the full potential of fourth-generation synchrotron sources in ultrafast materials dynamics.

\section{Conclusion}
Fast, bright, and radiation-hard scintillators are a key enabling technology for ultrafast hard X-ray imaging and diffraction at fourth-generation synchrotron sources such as the APS-U. In particular, MHz-rate dynamic materials experiments require scintillators with high X-ray stopping power, high light yield, and decay times substantially shorter than conventional LYSO/LSO in order to reduce inter-frame persistence and preserve temporal fidelity at frame intervals of 77 ns and below.

In this work, we evaluated Ce-doped lanthanum bromide and cerium bromide scintillators as leading candidates for indirect ultrafast X-ray detection. Compared with LYSO, commercial LaBr$_3$:Ce and CeBr$_3$ offer significantly faster decay components while maintaining comparable scintillation output, making them attractive for hard X-ray movies at APS-U-relevant time scales. Through laboratory prototype development, optical characterization, packaging studies, and initial ultrafast X-ray phase-contrast imaging tests at the Dynamic Compression Sector, we demonstrated that the major practical limitation of these materials---their hygroscopicity---can be substantially mitigated through appropriate handling, encapsulation, and optical-coupling strategies.

The results indicate that LaBr$_3$:Ce and CeBr$_3$ can provide a viable path toward brighter and faster scintillator screens for ultrafast dynamic materials experiments, provided that the complete detector chain is optimized. In particular, efficient light extraction, robust hermetic packaging, and high-quantum-efficiency conversion of near-UV/blue scintillation light remain essential for maximizing signal-to-noise ratio at MHz frame rates. These findings establish an experimental basis for deploying bromide scintillators in APS-U beamline instruments and motivate continued development of integrated scintillator--optics--camera assemblies.

Future work will focus on long-duration stability under realistic beamline conditions, radiation-damage assessment, optimization of scintillator thickness and optical coupling for specific imaging and diffraction geometries, and comparison with emerging alternatives such as perovskite and high-entropy scintillators. Together, these efforts will support the development of next-generation indirect detectors capable of capturing irreversible or single-impact ultrafast materials dynamics with improved temporal resolution, photon detection efficiency, and image quality.

\appendices
\section{An additional examples of X-ray imaging with a broken defective scintillator}
\begin{figure}[!hbt]
\centering
\includegraphics[width=3.6 in]{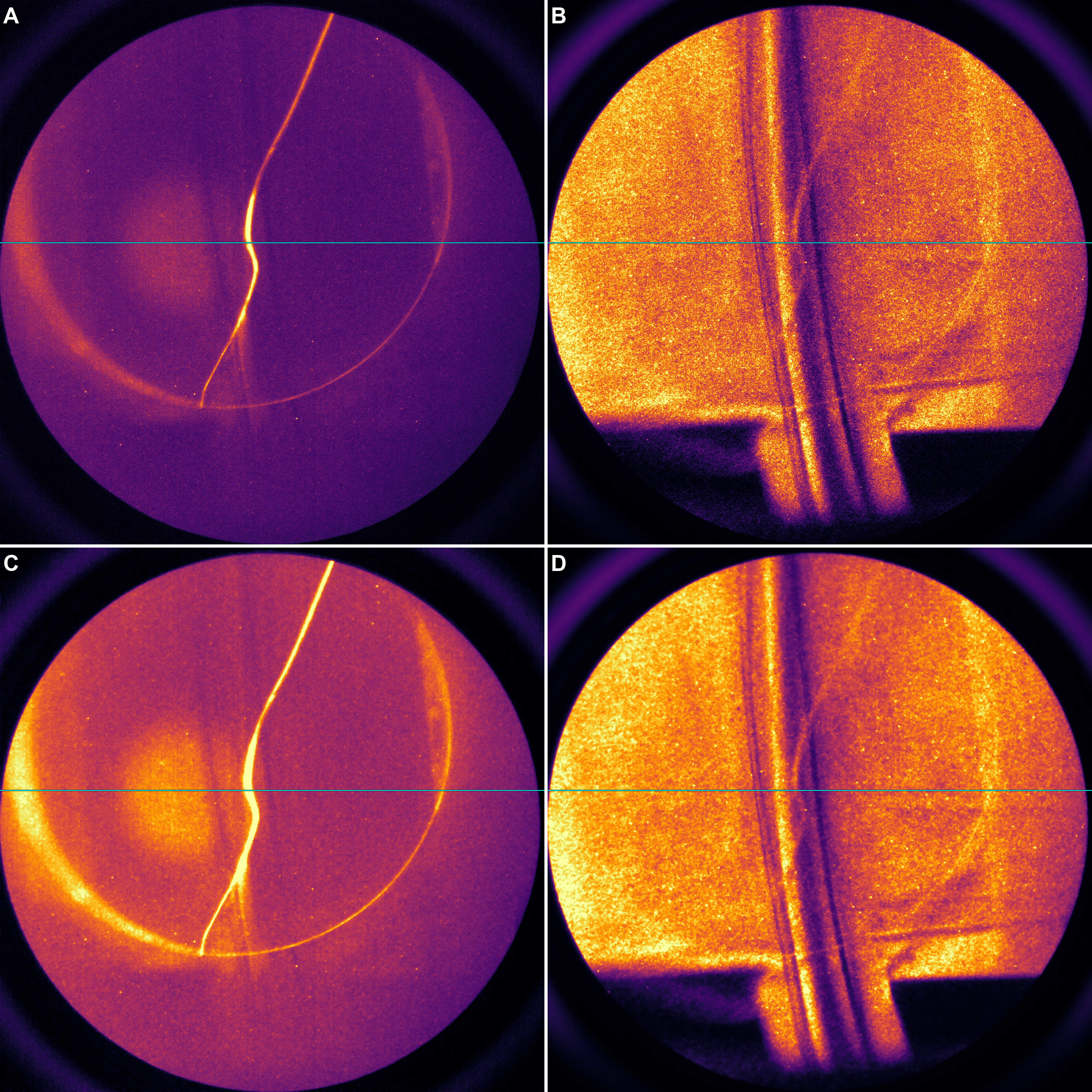}
\caption{Laboratory X-ray Imaging using Ag $k_\alpha$ and $k_\beta$ emission as the source and a 200-\textmu m thick LYSO scintillator as the X-ray-to-visible converter. A crack was observed across the scintillator crystal, which shows enhanced scintillator light emission along the crack and edges.}
\label{fig_broken:Ag_Vis_LYSO}
\end{figure}

\begin{figure}[!hbt]
\centering
%\fbox{\rule{0pt}{1.6in}\rule{0.95\linewidth}{0pt}}
\includegraphics[width=3.6 in]{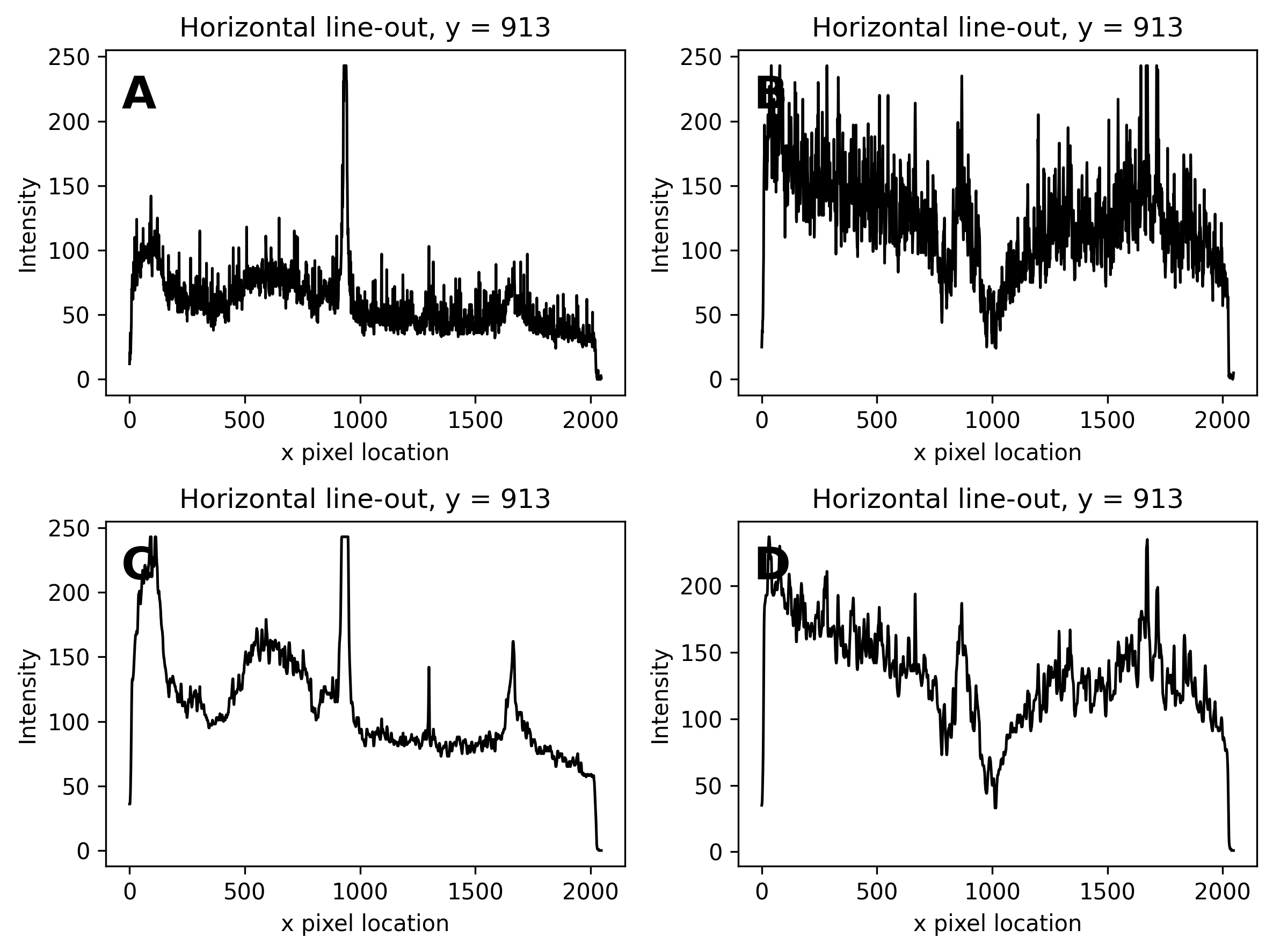}
\caption{Lineout analysis for the image pairs in Fig.~\ref{fig_broken:Ag_Vis_LYSO}.}
\label{fig_broken:Ag_Vis_LYSO_lo1}
\end{figure}

The thin scintillators and assemblies are fragile for various reasons. Fig.~\ref{fig_broken:Ag_Vis_LYSO} and Fig.~\ref{fig_broken:Ag_Vis_LYSO_lo1} correspond to Fig.~\ref{fig:Ag_Vis_LYSO} and Fig.~\ref{fig:Ag_Vis_LYSO_lo1} when the 200-\textmu m thick LYSO scintillator was broken near the middle across the full surface, showing the bright emission of scintillator light along the crack. The brightest region is near where the X-ray source is.

\section*{Acknowledgment}
Z. W. wishes to thank Washington State University staff for helping with target preparation, executing the dynamic experiments, logistics and safety at the Dynamic Compression Sector (DCS) beam line of APS-U. This work is supported in part by the LANL Office of Experimental Sciences (Dynamic Materials Properties and Advanced Diagnostics) and is
based upon work performed at the Dynamic Compression Sector, which is operated by
Washington State University under the U.S. Department of Energy (DOE)/National Nuclear
Security Administration award no. DE-NA0003957. This research used resources of the Advanced
Photon Source, a DOE Office of Science User Facility operated for the DOE Office of Science by
Argonne National Laboratory under contract no. DE-AC02-06CH11357.

\bibliographystyle{IEEEtran}
\bibliography{refs}

\end{document}